\documentclass[twocolumn,twoside,slac_two]{revtex4}
\usepackage{graphicx}
\usepackage{fancyhdr}
\begin{document}

\title{Novel Multi-pixel Silicon Photon Detectors and Applications in T2K}

%

\author{D. Beznosko (for T2K collaboration)}
\affiliation{Department of Physics and Astronomy, Stony Brook University, Stony Brook, NY, 11790-3800, USA}

\begin{abstract}
     Nowadays, numerous fields such as High Energy Physics (HEP), medical imaging devices, portable radiation detectors etc., require a robust, miniature, reliable and readily available photon detector that is stable in a variety of environments, such as the presence of strong magnetic fields. The recently available $\sim$1mm$^{\textrm{2}}$ active area Multi-pixel Photon Counter (MPPC) sensors, produced by Hamamatsu Photonics, have been found to be reliable and an attractive choice for the HEP applications.

     The following sensor characteristics have been thoroughly tested by T2K collaboration: gain, dark noise, detection efficiency, reliability. These appear to be stable; in addition, the characteristic spread between numerous devices was assessed. Sensors with larger area are being developed for imaging and direct-to-scintillator coupling purposes.
\end{abstract}

\maketitle

\thispagestyle{fancy}


\section{Introduction}
    T2K is the Tokai to Kamioka Experiment, a long baseline neutrino oscillation experiment based in Japan.  It consists of three parts; an artificial neutrino beam line at the J-PARC facility, a Near Detector suite 280 meters from the beam start point (ND280)~~\cite{nd280}, and Super-Kamiokande acting as far detector, 295 km from Tokai.  Among it's main physics goals is a measurement of the $\theta_{\textrm{13}}$ neutrino oscillation parameter, by searching for $\nu_{\textrm{e}}$ appearance in the $\nu_{{\mu}}$ beam.

     The beam line consists of the 30 GeV proton beam incident on graphite target. This is followed by focusing horn system, helium filled decay volume, beam dump, muon monitor and ND280.  The neutrino off-axis beam energy at ND280 off-axis detector is expected at ~0.6GeV at peak intensity value with a narrow energy band.

     ND280 on-axis detector is to measure beam parameters such as direction, profile and intensity. ND280 off-axis detector consists of several sub-detectors, all in $\sim$0.2T magnetic field for accurate momentum measurements and charge discrimination. The intended measurements to be made using this detector group include neutrino beam energy spectrum, flux, $\nu_{\textrm{e}}$ contamination, background, and cross sections measurements.
\begin{figure}[h]
\centering
\includegraphics[width=80mm]{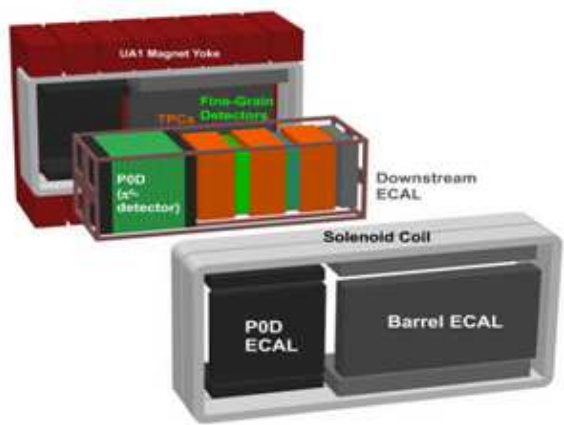}
\caption{The ND280 off-axis detector.} \label{nd280_figure}
\end{figure}

The following are the off-axis sub-detectors as positioned inside UA1 magnet (Figure~\ref{nd280_figure}):
\begin{itemize}
\item P0D (Pi-Zero Detector),
\item FGD (fine grained detector),
\item TPC (time projection chamber),
\item ECAL (electromagnetic calorimeter),
\item SMRD (Side Muon Range Detector, inside magnet yoke).
\end{itemize}
    With exception of TCP, each detector utilizes scintillating bars, wavelength shifting fibers and a photon detector with some readout/digitization scheme. Therefore, a use of efficient, high performance photon detector that operates in magnetic field is required.
\section{Photodetectors Choice}
\subsection{P0D as Off-axis Sub-detector}
     The Pi-Zero Detector (P0D) is among the off-axis sub-detectors. It is primarily designed to measure Neutral Current $\pi^{\textrm{0}}$ production cross section on water to reduce the background. P0D is a series of alternating x-y orientation of the TiO$_{\textrm{2}}$ coated triangular extruded scintillator bars with co-extruded hole~\cite{scintillator}, with Kuraray~\cite{kuraray} Y11 wavelength shifting fibers and single side readout.

     For the readout, following two types of photodetectors were tested:
\begin{itemize}
\item MCPMAPMT - Micro-channel Plate Multi-anode PMT~\cite{burle}
\item MPPC - Multi-Pixel Photon Counter~\cite{hamamatsu}.
\end{itemize}
\subsection{MCPMAPMT Description and Testing}
     A Burle~\cite{burle} 85011-501 device was used for the performance testing in the magnetic field. It is based on multichannel plate technology, with typical gain of about 7*10$^{\textrm{7}}$ and anode uniformity of 1:1.5. For nominal operations, the high voltage bias (HV) of -2600V is recommended by the manufacturer. Maximum photon detection efficiency (PDE) is achieved for light at 400nm wavelength (the peak output of Y11 WLS fiber is $\sim$500nm). 

\begin{figure}[h]
\centering
\includegraphics[width=80mm]{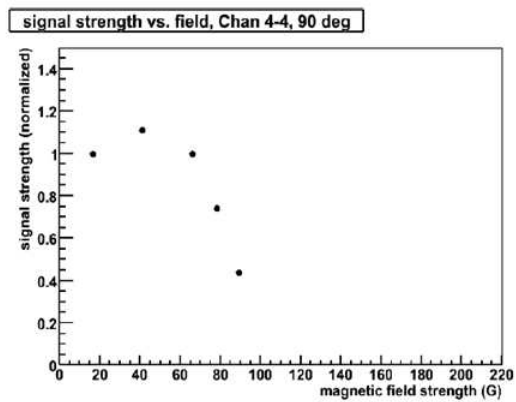}
\caption{Burle 85011-501 MCPMAPMT output vs. magnetic field strength at 90$^{\textrm{o}}$ between device axis and the field.} \label{mapt_testing}
\end{figure}

     This device was tested in the magnetic field presence~\cite{lisa}. A strong dependence of the output to the angle between the device axis and magnetic field was found. Figure~\ref{mapt_testing} shows the output dependence on the field strength at the 90$^{\textrm{o}}$ between device 
    
	In addition, cross-channel crosstalk was determined to be, on average, about $\sim$9-10$\%$ between nearest neighbor anodes that are in 8x8 square arrangements.

    Thus, this device was found to perform poorly in the magnetic field environment. An alternative had to be found and tested.
\subsection{MPPC Description}
\subsubsection{Geiger mode introduction}
     Geiger mode is a runaway avalanche in Avalanche Photo Diode (APD). In this mode, detector will only indicate photon detection, not light amount. However, the APD gain can be drastically increased. To avoid the avalanche becoming self-sustaining, a limiting circuitry is introduced for an effect that is generally called quenching. Two types are defined: active and passive quenching.
	 
     Active quenching constituted an external circuit, typically a fast controller chip that cuts bias voltage to quickly stop the avalanche after its development is detected by the steep increase in current drawn by the photon detector. This is a fast and effective method, but it requires additional circuitry that normally is not on the photosensor itself.
	 
     Passive quenching typically is an internal to sensor high value resistor. As current drawn by the sensor will rise, the voltage drop over the resistor will reduce the total bias at the photosensor, thus quenching the avalanche. This method is slower than active quenching and produces an output with a longer, exponentially decaying, tail, but is cheaper and can be built into the device during the manufacturing process.

\begin{figure}[h]
\centering
\includegraphics[width=80mm]{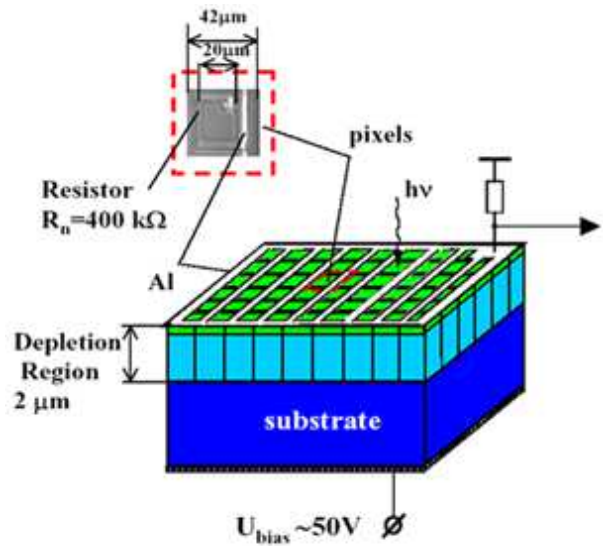}
\caption{Multi-pixel silicon family detector schematic~\cite{bondarenko}.} \label{sipm_schematic}
\end{figure}

     A typical multi-pixel silicon family photon detector has from ~100 to ~1600 pixels (1mm$^{\textrm{2}}$ models). Each pixel (typ. up to 100$\mu$m x 100$\mu$m) is an APD in limited Geiger mode, with passive quenching by surface film resistor (Figure~\ref{sipm_schematic}). Since each pixel will avalanche with a single photon, then for low light levels with the single pixel occupancy (photon$\#$/pixel$\#$) less than 1, the total output of such multi-pixel detector will be proportional to incident light flux.
     An incident photon, if absorbed, produces electron-hole pair. Accelerated by intense electric field within the pixel (typ. up to 10$^{\textrm{6}}$ V/m), either electron or hole, depending on the positive or negative biasing, can start avalanche in a pixel (pixel is said to 'fire').
     Each pixel is passively quenched by means of a film resistor that is typically located on the device surface. As avalanche develops, rising current causes a voltage drop over resistor and reduces pixel bias below avalanche limit. This additionally protects the sensor from damage from high light levels.
     The dark or uncorrelated, noise of this device is due to the thermally produced electron-hole pair can cause firing in the pixel. It has same output shape as from a detected single photon and is indistinguishable from it. In addition, there is a correlated noise produced by inter-pixel cross-talk within a single detector, and afterpulsing. Both will be discussed later.
\subsubsection{MPPC Basic Characteristics}
\begin{figure}[h]
\centering
\includegraphics[width=80mm]{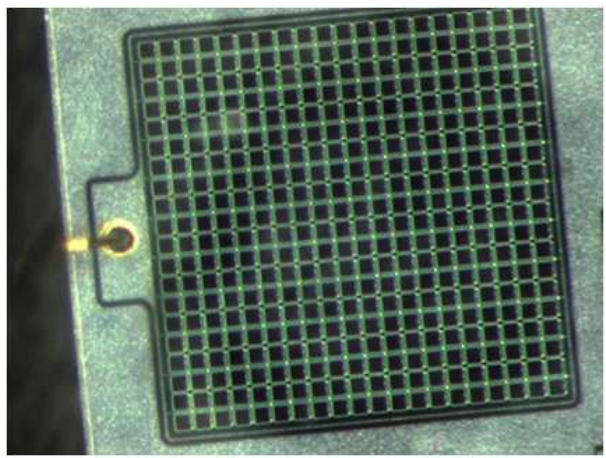}
\caption{S10361-050U MPPC magnified.} \label{mppc_microscope}
\end{figure}

     MPPC, or Multi-pixel Photon Counter, is one of the latest developments in the multi-pixel detector family by Hamamatsu. The following characteristics at 25$^{\textrm{o}}$C of the MPPC photodiode are provided by the manufacturer~\cite{hamamatsu}: 
     The main characteristics from manufacturer for the 400\footnote{a sub-model S10363-050U was produced for T2K with  667 pixels, 1.3x1.3mm2 active area} pixel S10361-050U model are given in Table~\ref{mppc_char}, and its photograph under a microscope is presented in Figure~\ref{mppc_microscope}.
\begin{table}[h]
\begin{center}
\caption{Characteristics for S10361-050U MPPC.}
\begin{tabular}{|l|c|c|c|}
\hline \textbf{Name} & \textbf{Value}
\\
\hline Number of pixels & 400\\
\hline Chip size & 1.5 x 1.5 mm$^{\textrm{2}}$\\
\hline Active area & 1 x 1 mm$^{\textrm{2}}$\\
\hline Pixel size & 50 x 50 $\mu$m$^{\textrm{2}}$\\
\hline Pixel effective size & 38.1 x 38.8 $\mu$m$^{\textrm{2}}$ \\
\hline Geometric efficiency  & 61.5 $\%$\\
\hline Time resolution & 220 ps \\
\hline Temp. coeff. of  & 50 mV/$^{\textrm{o}}$C\\
breakdown voltage &\\
\hline Dark count rate & $\sim$5$\cdot$10$^{\textrm{5}}$/sec\\
\hline Gain typ. & 7$\cdot$10$^{\textrm{5}}$\\
\hline Operating voltage & $\sim$70V\\
(varies from sensor to sensor)&\\
\hline
\end{tabular}
\label{mppc_char}
\end{center}
\end{table}

\section{3.	MPPC Characterization and Testing}
\subsection{Output PE Separation}
\begin{figure}[h]
\centering
\includegraphics[width=80mm]{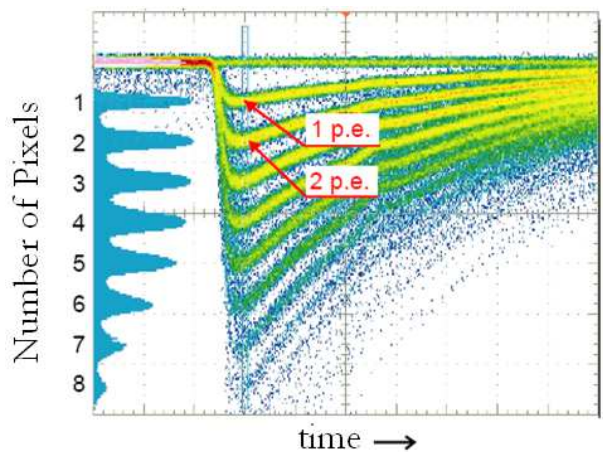}
\caption{MPPC output PE spectrum~\cite{mppc_PE_scope}.} \label{mppc_PE_scope}
\end{figure}

\begin{figure}[h]
\centering
\includegraphics[width=80mm]{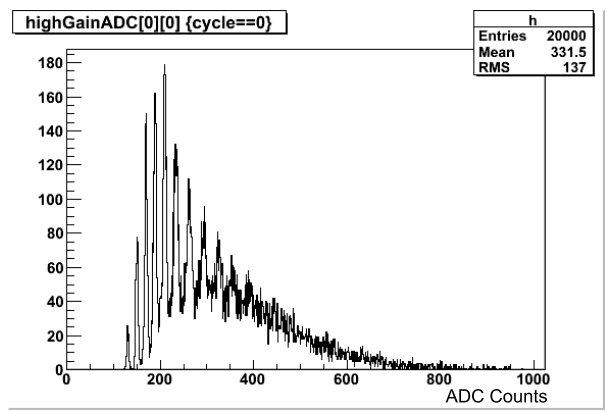}
\caption{MPPC PE output using ADC.} \label{mppc_adc}
\end{figure}

     Since MPPC is a multi-pixel device, with each pixel providing approximately equal output signal, the combined output of the MPPC exhibits a Photo Electron (PE) structure, such as in oscilloscope trace in Figure~\ref{mppc_PE_scope}~\cite{mppc_PE_scope}.
     Each line is traced when single pixel, or two or three fire at the same time. All testing is done at room temperature ($\sim$25$^{\textrm{o}}$C) unless stated otherwise. A similar spectrum measured by Analog-to-Digital Converter (ADC) is in Figure~\ref{mppc_adc}. 
     Figure~\ref{mppc_adc} was obtained by flashing LED at low light level directly onto MPPC surface. The first peak is the pedestal, the second peak is a single fired pixel, with each subsequent peak representing an additional pixel fired. Here, several peaks are still clearly discernable, thus this photodetector feature can be used for the convenient calibration of the MPPC output.

\subsection{Cross-talk, Afterpulsing and Dark Noise}
     MPPC output has the following features that arise from its multi-pixel structure and avalanche amplification scheme. These are cross-talk between pixels and the afterpulse of the pixel.
     Cross-talk occurs when, from a firing from signal detection or dark noise pixel, an electron or recombination photon reaches some neighboring pixel and causes it to fire effectively simultaneously. The probability of cross-talk acts as additional amplification is a sense, however, needs to be understood for accurate energy output calibration.
\begin{figure}[h]
\centering
\includegraphics[width=80mm]{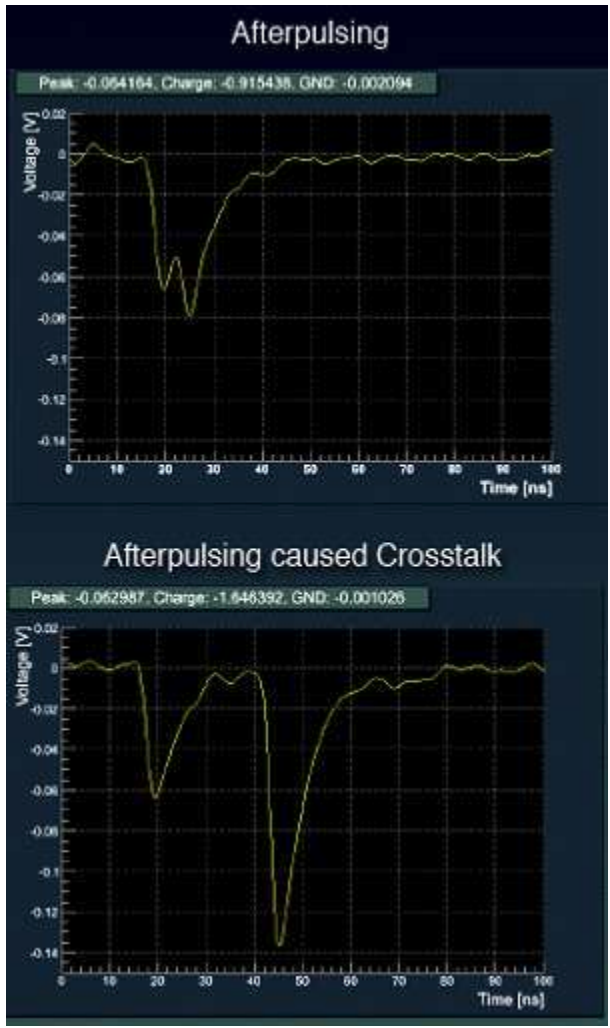}
\caption{(top): afterpulsing effect. (bottom): crosstalk cause by afterpulse.} \label{crosstalk_afterpulse1}
\end{figure}
     Afterpulse is the same pixel firing shortly after it had fired due to signal detected, dark noise or cross-talk. In majority, this is due to the electrons that get trapped on impurities and structure defects in the pixel structure and can cause an avalanche later. Typical lifetime of such trap states is from several ns to a couple hundreds of ns. A pixel needs a recovery time after firing so that it can support full avalanche. If afterpulse occurs before that time, the output will be of partial amplitude as in Figure~\ref{crosstalk_afterpulse1}(top) \cite{crosstalk_afterpulse1}. Afterpulse, in turn, can cause crosstalk. Figure~\ref{crosstalk_afterpulse1}(bottom) \cite{crosstalk_afterpulse1} shows when an afterpulse causes crosstalk increasing its amplitude.
\begin{figure}[h]
\centering
\includegraphics[width=80mm]{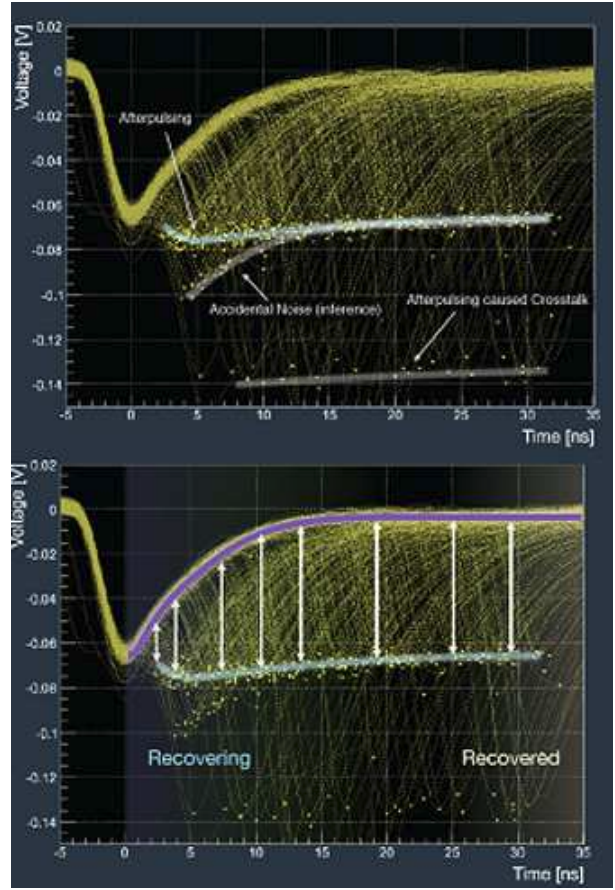}
\caption{(top): Scope trace shows MPPC output structure. (bottom): Afterpulse amplitude as function of recovery time.} \label{crosstalk_afterpulse2}
\end{figure}

     Figure~\ref{crosstalk_afterpulse2}(top) \cite{crosstalk_afterpulse1} shows the full trace of the MPPC dark noise output. The afterpulse and cross-talk are clearly visible, also one can notice the amplitude change in afterpulses (Figure~\ref{crosstalk_afterpulse2}(bottom) \cite{crosstalk_afterpulse1}). These can be fit  with double exponent to get two recovery times~\cite{crosstalk_afterpulse1}:
\begin{itemize}
\item Short time const - $\sim$19ns
\item Long time const - $\sim$85ns.
\end{itemize}
After-pulsing probability can also be obtained:
\begin{itemize}
\item Short - $\sim$6.1$\%$
\item (long) - $\sim$5.9$\%$
\end{itemize}
with cross-talk probability of $\sim$5.7$\%$.

\begin{figure*}[t]
\centering
\includegraphics[width=135mm]{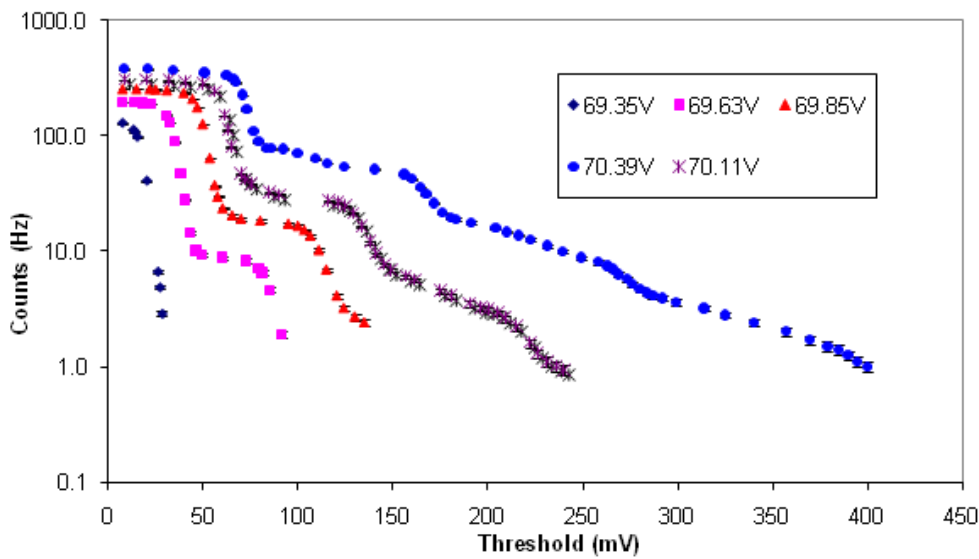}
\caption{MPPC noise rate vs. biasing voltage and discriminator threshold.} \label{steps}
\end{figure*}

\begin{figure}[h]
\centering
\includegraphics[width=80mm]{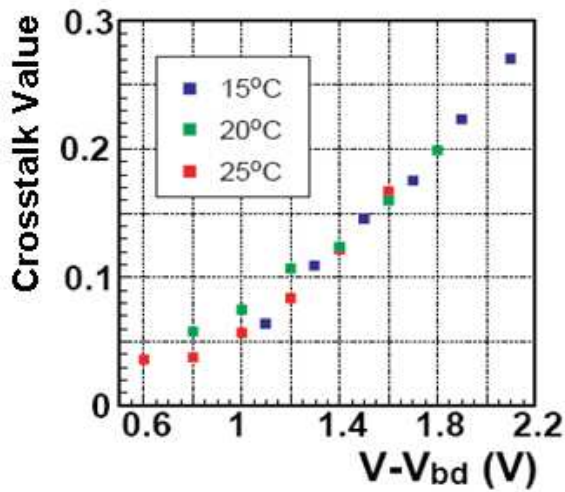}
\caption{Cross-talk rate dependance on biasing voltage and temperature.} \label{crosstalk_to_bias_T}
\end{figure}

Temperature and bias dependence of cross-talk rate was studied. The temperature seems to be a low factor with effect dominated by biasing voltage above a breakdown (i.e. voltage at which an avalanche can be sustained in the pixel, denoted VB or V$_{\textrm{bd}}$). Figure~\ref{crosstalk_to_bias_T}~\cite{colors} shows the result of that study.

     With the above, a simple and more visual way for cross-talk and afterpulse check in presented in Figure~\ref{steps}. $\sim$100x amplifier was used. This is a graph of average dark noise frequency (averaged over a minute at room temperature) for different biases at different threshold values of the discriminator. Each plateau length is the amplitude of single PE, the slope at the ends of plateaus shows the effects of afterpulsing (otherwise it would be more vertical, with a slope due mostly to pixel to pixel output variations). Ratios between frequencies of each plateau give the approximate rate of the cross-talk. Note that for some high noise sensors ($\sim$1MHz), the accidental coincidence of different pixels with dark noise may be significant.

\begin{figure}[h]
\centering
\includegraphics[width=80mm]{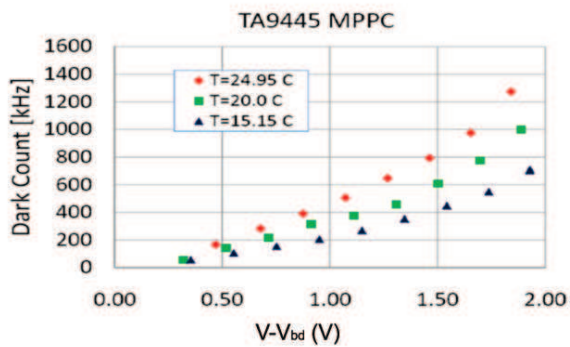}
\caption{Dark noise vs. biasing voltage and temperature.} \label{dark_noise_vs_bias_and_temp}
\end{figure}

In addition, the dependence of the dark noise on biasing voltage and temperature was measured at threshold of 0.5 PE. The result is presented in Figure~\ref{dark_noise_vs_bias_and_temp}~\cite{colors}.
\subsection{MPPC Gain and PDE}
The photosensor gain and Photon Detection Efficiency (PDE) dependence on temperate and biasing voltage were studied.

\begin{figure}[h]
\centering
\includegraphics[width=80mm]{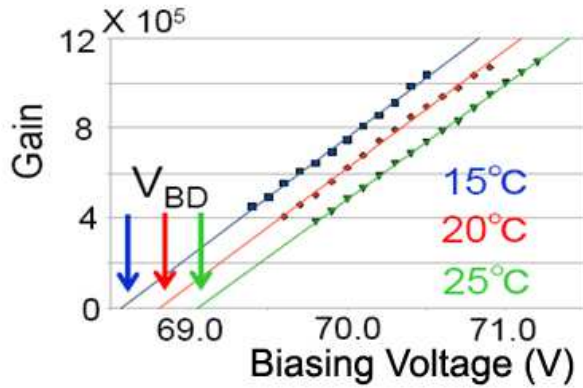}
\caption{MPPC gain vs. biasing voltage and temperature.} \label{gain_bias_temp}
\end{figure}

The dependence of gain on bias and temperature is shown in Figure~\ref{gain_bias_temp}~\cite{colors}. The temperature coefficient of breakdown voltage change with temperature value used was 50mV/$^{\textrm{o}}$C.

\begin{figure}[h]
\centering
\includegraphics[width=80mm]{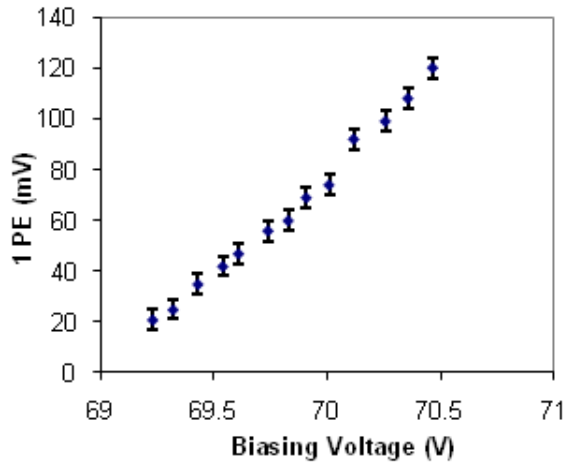}
\caption{MPPC single PE amplitude vs. biasing voltage.} \label{pe_amplitude_bias}
\end{figure}

Amplitude of a single PE, using ~100x amplifier, was measured for various biasing voltages (Figure~\ref{pe_amplitude_bias}).

\begin{figure}[h]
\centering
\includegraphics[width=80mm]{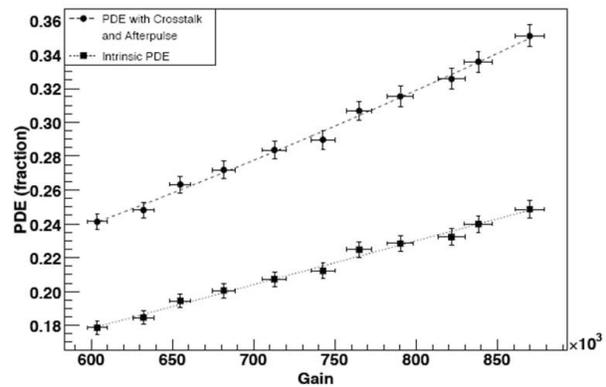}
\caption{Apparent and intrinsic PDE vs. MPPC gain.} \label{PDE}
\end{figure}

 Figure~\ref{PDE}~\cite{colors} presents the PDE dependence of the MPPC sensor vs. gain at 470nm (that is linear with bias voltage). The bottom and top lines in the figure illustrate the difference of the apparent from intrinsic PDE caused by the effects of cross-talk and afterpulsing. Thus, afterpulse and cross-talk act to increase the apparent sensor output. No noticeable change of PDE with temperate has been observed \cite{colors}.

\begin{figure}[h]
\centering
\includegraphics[width=80mm]{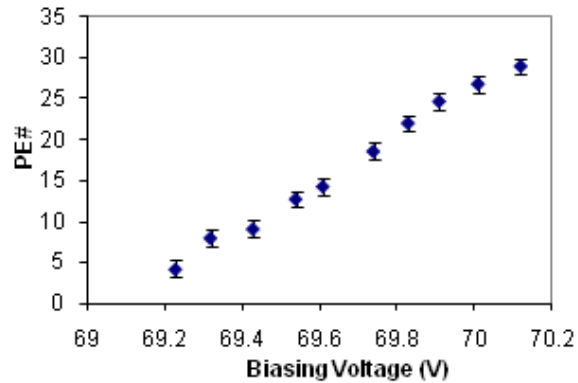}
\caption{Signal amplitude in PE vs. biasing voltage.} \label{PE_out}
\end{figure}

With biasing voltage, the dark noise rate grows, so it's not practical to use these sensors at the upper bound of the operating range. In addition, as biasing voltage increases, the PDE, cross-talk and afterpulse increase as well (Figure \ref{PE_out}), causing increase in the number of PE seen in the output for the same amount of incident light.

\begin{figure*}[t]
\centering
\includegraphics[width=135mm]{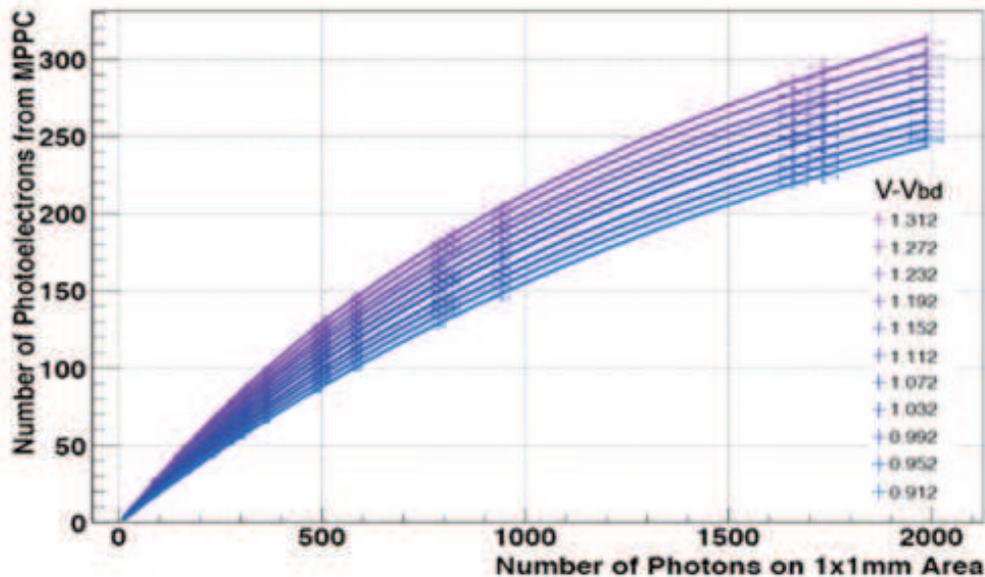}
\caption{MPPC output vs. applied signal for a range of biasing voltages.} \label{linear}
\end{figure*}

The MPPC has a limited number of pixels. As it is assumed that the incoming photons spread equally over sensor surface, and each pixel's output doesn't increase if more than one photon is detected at the same time (or before pixel recovers). This non-linearity of the output to the applied signal vs. bias is illustrated in Figure \ref{linear} \cite{colors} (for 400 pixel device).

\subsection{MPPC Large Scale Deployment}
For P0D, $\sim$11k MPPC were ordered and tested for P0D. This testing involved obtaining ADC spectra using LED before installation, dark noise after installation, radioactive source signal during each module scanning, and dark noise testing after shipping detector to Japan. During initial testing, approximately 14 sensors were found to be bad, 2 with damaged surface. Four more sensors were replaced for various reasons after testing in Japan.

\begin{figure}[h]
\centering
\includegraphics[width=80mm]{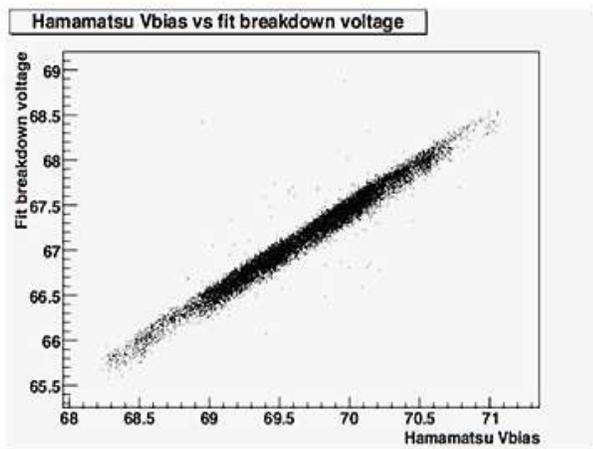}
\caption{MPPC operating voltages spread for $\sim$11k sensors.} \label{spread}
\end{figure}

Figure \ref{spread} is the spread of operating voltage between all P0D sensors \cite{csu}. The voltage values used are the manufacturer recommended ones.

As an additional test of the detector, cosmic ray muons were detected once in Japan.  This first real-world use of the P0D demonstrated that the devices worked correctly, with an average PE output from each scintillator bar ($\sim$17mm active material, averaged over bar length) was $\sim$20PE/MeV. Similarly high light yeild is reported by other sub-detector groups \cite{SMRD} \cite{other_detectors}.

\subsection{Multi-pixel Family Detector Testing}
Besides tests done on MPPC, rigorous testing of similar sensors from other manufacturers was performed both inside and outside of T2K project. These included 1MRad irradiation, alignment with fiber and performance in 9T magnetic field \cite{dima}.

\section{Conclusion}

Overall, MPPC sensors are found to be reliable and an attractive choice for HEP applications, especially in strong magnetic field environments. Sensor characteristics such as gain, dark noise, PDE appear to be stable over time and within magnetic field. The real-life sensor performance during P0D initial testing was observed. A low operating voltage spread among $\sim$11k devices was measured. A promising long term stability and currently observed low failure rate are to be verified from ND280 lasting performance.

Currently, sensors with larger area are being developed. A range of other applications for these sensors have already been identified, including use in medical imaging devices, portable radiation detectors and security systems.

\bigskip 

\begin{thebibliography}{99}
%
\bibitem{nd280}``T2K ND280 Conceptual Design Report'',  T2K Internal Document.
See also D.~Karlen, Nucl. Phys. B (Proc. Suppl.) 159 (2006) 91;
Yu.Kudenko, Nucl. Instr. Meth. A598 (2009) 289, arXiv:0805.0411.
[physics.ins-det].
\bibitem{scintillator} D. Beznosko, A. Bross, A. Dyshkant, A. Pla-Dalmau V. Rykalin, "FNAL-NICADD Extruder", FERMILAB-PUB-05-344, Jul 29, 2005
\bibitem{kuraray} Kuraray America Inc. 200 Park Ave, NY 10166,USA;
\bibitem{burle} BURLE INDUSTRIES, INC., 1000 New Holland Avenue, Lancaster, Pennsylvania 17601-5688 U.S.A.;   http://www.burle.com
\bibitem{hamamatsu}Hamamatsu Photonics, 314-5,Shimokanzo, Toyooka-village, Iwatagun,Shizuoka-ken, 438-0193 Japan;
http://www.hamamatsu.com;

M.~Yokoyama  et al., arXiv:0807.3145[physics.ins-det].

\bibitem{lisa} L. Whitehead, K. Kobayashi, C. Yanagisawa, "Test of the Burle Micro-Channel Plate Multi-Anode Photomultiplier Tube", April 19, 2006; http://www.phy.bnl.gov/~whitehd/MCPnote.pdf

\bibitem{bondarenko} G. Bondarenko et al., Nucl. Instr. and Meth., A242 (2000) p.187

\bibitem{mppc_PE_scope} MPPC product information datasheet "MPPC (Multi-Pixel Photon Counter) S10362 series"

\bibitem{crosstalk_afterpulse1} "Study of Afterpulsing of MPPC with Waveform Analysis", Hideyuki Oide et al.

\bibitem{colors} ] "Development of Multi-Pixel Photon Counters", M. Yokoyama et al., SNIC Symposium, Stanford, CA 3-6 April, 2006

\bibitem{csu} "MPPC Testing at CSU", D. Ruterbories et al., T2K Workshop, Jan 2009.

\bibitem{SMRD} A.Izmaylov et al. "Scintillator counters with WLS fiber/MPPC readout for the side muon range detector (SMRD)of the T2K experiment" arXiv:0904.4545

\bibitem{other_detectors} "Using MPPCs for T2K near detector", F.Retiere, PD09 conference.; http://www-conf.kek.jp/PD09/

\bibitem{dima} D. Beznosko, G. Blazey, A. Dyshkant, V. Rykalin, V. Zutshi, "Effects of the Strong Magnetic Field on LED, Extruded Scintillator and MRS Photodiode", NIM A 553 (2005) 438-447



\end{thebibliography}

\end{document}